Article type: Article

# Ultrafast Epitaxial Growth of Metre-Sized Single-Crystal Graphene on Industrial Cu Foil

Xiaozhi Xu[a,b#], Zhihong Zhang[a,b#], Jichen Dong[c], Ding Yi[c], Jingjing Niu[a], Muhong Wu[a], Li Lin[d], Rongkang Yin[e], Mingqiang Li[a], Jingyuan Zhou[d], Shaoxin Wang[a], Junliang Sun[f], Xiaojie Duan[d,e], Peng Gao[a,d,g], Ying Jiang[g,h], Xiaosong Wu[a,g], Hailin Peng[d], Rodney S. Ruoff[c,i], Zhongfan Liu[d], Dapeng Yu[a,g,j], Enge Wang[g,h], Feng Ding[c,k], Kaihui Liu[a,d,g]

[a] State Key Laboratory for Mesoscopic Physics, School of Physics, Peking University, Beijing 100871, China

[b] Academy for Advanced Interdisciplinary Studies, Peking University, Beijing 100871, China

[c] Center for Multidimensional Carbon Materials (CMCM), Institute for Basic Science (IBS), Ulsan 689-798, Republic of Korea

[d] Centre for Nanochemistry, College of Chemistry and Molecular Engineering, Peking University, Beijing 100871, China

[e] Department of Biomedical Engineering, College of Engineering, Peking University, Beijing 100871, China

[f] College of Chemistry and Molecular Engineering, Peking University, Beijing 100871, China

[g] Collaborative Innovation Centre of Quantum Matter, Beijing 100871, China

[h] International Centre for Quantum Materials, Peking University, Beijing 100871, China

[i] Department of Chemistry, Ulsan National Institute of Science and Technology (UNIST), Ulsan 689-798, Republic of Korea

[j] Department of Physics, South University of Science and Technology of China, Shenzhen 518055, China

[k] School of Materials Science and Engineering, Ulsan National Institute of Science and Technology (UNIST), Ulsan 689-798, Republic of Korea

\# These authors contributed equally to this work

\* Correspondence: khliu@pku.edu.cn; f.ding@unist.ac.kr; egwang@pku.edu.cn

**ABSTRACT**

A foundation of the modern technology that uses single-crystal silicon has been the growth of high-quality single-crystal Si ingots with diameters up to 12 inches or larger. For many applications of graphene, large-area high-quality (ideally of single-crystal) material will be enabling. Since the first growth on copper foil a decade ago, inch-sized single-crystal graphene has been achieved. We present here the growth, in 20 minutes, of a graphene film of $5 \times 50$ cm$^2$ dimension with > 99% ultra-highly oriented grains. This growth was achieved by: (i) synthesis of sub-metre-sized single-crystal Cu(111) foil as substrate; (ii) epitaxial growth of graphene islands on the Cu(111) surface; (iii) seamless merging of such graphene islands into a graphene film with high single crystallinity and (iv) the ultrafast growth of graphene film. These achievements were realized by a temperature-driven annealing technique to produce single-crystal Cu(111) from industrial polycrystalline Cu foil and the marvellous effects of a continuous oxygen supply from an adjacent oxide. The as-synthesized graphene film, with very few misoriented grains (if any), has a mobility up to ~ 23,000 cm$^2$V$^{-1}$s$^{-1}$ at 4 K and room temperature sheet resistance of ~ 230 Ω/□. It is very likely that this approach can be scaled up to achieve exceptionally large and high-quality graphene films with single crystallinity, and thus realize various industrial-level applications at a low cost.

1. Introduction

Given the many possibilities for applications of high-quality graphene[1, 2], routes towards its industrial-scale synthesis gradually are of great importance[3-5]. For most industrial-level devices (and other) applications, large single-crystal graphene (SCG) films are ideal for top-down processing. In the past 10 years, the SCG island size has increased more than four orders of magnitude, from micrometres to inches[6-25].

There are two routes towards large SCG—(**a**) to allow only one graphene nucleus to grow into a large SCG film[9-18], or (**b**) to 'seamlessly stitch' multiple aligned graphene islands into an SCG film[19-25]. Recently, the synthesis of a 1.5-inch SCG was enabled by the local gas feeding technique, in which one nucleus only, grows[15]. SCG by route (**a**) usually requires a very long time for the one nucleus to grow by addition of carbon at the edges, while route (**b**) can in principle rapidly yield a large SCG film and thus looks favourable for scaled production. We identified four challenges for route (**b**): (*i*) a large-area single-crystal substrate with proper symmetry, such as the (111) surface of a fcc crystal or the Ge(110) surface[21-25]; (*ii*) epitaxial growth of aligned graphene islands; (*iii*) seamless stitching of a very large amount of the well aligned graphene islands into a continuous SCG film and (*iv*) fast growth of every large SCG island. Previous studies stated that challenges (*ii*) and (*iii*) had been met by graphene grown on Ge(110) or Cu(111) surfaces, but it was reported that there remained 2-5% unaligned graphene islands, which may degrade the graphene film's quality[22, 23]. Our recent study met challenge (*iv*) through the introduction of a continuous supply of oxygen from an adjacent oxide during the growth[17]. Therefore, the two challenges that we particularly address here are to achieve a proper type of large-area and easily available single-crystal substrate, and to improve the degree of alignment of the graphene islands.

Since both Cu(111) surface and graphene have $C_3$ rotation symmetry and their lattice mismatch is as small as 4%, super-large single-crystal Cu(111) foil is ideal for large-size SCG growth. However, but all metre-scale Cu foils in the market are polycrystalline (to the best of our knowledge) and available Cu(111) single crystals are mainly of inch size and of very high price (preventing its industrial applications). It was observed that proper thermal annealing will generate Cu(111) islands with size up to centimetre scale[18, 22], but further increasing the island size to metre scale is very challenging, probably due to the lack of driving forces to form a uniform single crystal. Also just as graphene growth, during the Cu annealing process, many nucleation centres will appear and form Cu(111) domains with different in-plane rotations. Although each domain can be of centimetre size, the large-scale Cu foil is still of polycrystal (Fig. S1 and Movie S1, online). It seems that the difficulty in producing super-large single-crystal Cu foil is as challenging as that in graphene.

**2. Experimental**

*2.1 Annealing polycrystalline Cu foil into single-crystal Cu(111) foil*

The industrial Cu foil (25 μm thick, 99.8 %, Sichuan Oriental Stars Trading Co. Ltd.) was placed on a quartz substrate and then loaded into a CVD furnace (Tianjin Kaiheng Co. Ltd., customer designed). The central region of the furnace was kept at 1030 °C with 500 sccm Ar when Cu foil was continuously sliding through the furnace centre.

*2.2 Growth of metre-sized graphene film*

The single-crystal Cu(111) foil continuously passed through the furnace tube very close to a quartz substrate at the speed of 2.5 cm/min. For the growth, the CVD system was heated to 1030 ℃ under 500 sccm Ar and 10 sccm $H_2$. $CH_4$ (1-5 sccm) was introduced during the growth. After growth, the system was cooled down naturally under 500 sccm Ar and 10 sccm $H_2$.

*2.3 $H_2$ etching and 'Ultraviolet oxidization' of graphene on Cu*

After the growth of graphene, the $CH_4$ was cut off and the sample was exposed to 10 sccm $H_2$ for 15 min at the growth temperature to make the $H_2$ etching occur, with flow of 500 sccm Ar.

A single-crystal Cu(111) foil covered with aligned graphene islands and a polycrystalline Cu foil with unaligned graphene islands were placed side by side in a UV-ozone cleaner (SAMCO's UV-1$^{TM}$, hot cathode, low-pressure mercury vapour lamp with active wavelengths of ~85 % 254 nm and ~15 % 185 nm ). Humid air was introduced into the chamber by connecting it to a water bubbler. The graphene/Cu substrate was irradiated with ultraviolet light under 500 sccm $O_2$ for 20 min to find out whether defective boundaries of the samples would be oxidized.

*2.4 Graphene transfer*

The as-grown graphene film was transferred onto $SiO_2$/Si or h-BN substrates by the PMMA-assisted method. The graphene film was spin-coated with PMMA and baked at 150 $^o$C for 5 min. Then, a 1 M $Na_2S_2O_8$ solution was used to etch the copper away. After being rinsed by deionized water, the PMMA/graphene was subsequently washed by isopropanol and then dried in air for 12 hours before it was placed onto the $SiO_2$/Si or h-BN substrate. Subsequently, the PMMA was dissolved by acetone.

*2.5 Graphene electrical device fabrication and electrical measurements*

Four- or six-terminal graphene Hall bars were fabricated using the standard electron-beam lithography (EBL) followed by $O_2$ plasma etching. The Pd/Au contact electrodes were patterned via EBL, electron-beam metal evaporation and lifting-off processes. The electrical measurements were carried out in an Oxford cryostat instrument with a variable temperature (1.5–300 K) insert. The magnetic field was applied perpendicular to the sample surface. Electrical signals were measured via low-frequency lock-in techniques.

The room temperature sheet resistance of the graphene films was measured by using a four-probe resistance measuring meter (CDE RESMAP 178) based on the four-point probe method to eliminate contact resistance. Each data is obtained from a 1×1 mm$^2$ area.

*2.6 Characterization*

Raman spectra were obtained with a LabRAM HR800 system with laser excitation wavelength of 532 nm and power of ~ 1 mW. Optical images were obtained with an Olympus BX51 microscope.

LEED measurements were performed using Omicron LEED system in UHV with base pressure $<3\times10^{-7}$ Pa. Gaussian fitting was used to determine the position of each single diffraction point of the LEED patterns. And the lattice orientation was determined by the least squares fitting method.

STM experiments were performed with a combined nc-AFM/STM system (Createc, Germany) at 5 K with base pressure $<7\times10^{-9}$ Pa. All of the STM topographic images were obtained in constant-current mode.

**3. Results and discussion**

Fig. 1a schematically illustrates our experimental design for the continuous production of single-crystal Cu(111) foil. In order to achieve a large single-crystal Cu(111) foil, we slowly slide a polycrystalline Cu foil through a hot zone (the central temperature is 1030 ℃, ~ 55 ℃ lower than the melting point of Cu) near the centre of a quartz tube (Fig. 1a). The temperature gradient around the central hot zone provides a driving force for the continuous motion of grain boundaries (GBs) in the Cu foil. This idea is similar to the traditional "Czochralski method", in which the temperature gradient at the interface between liquid and solid acts as the driving force for the growth of single-crystal silicon ingot. During single-crystal foil generation, one edge of the Cu foil was tapered, which ensured nucleation of only one Cu(111) grain at the very tip (Fig. 1b-e). Sliding the foil through the central hot zone was found to drive movement of the grain boundaries between the single-crystal and the polycrystalline regions and it was found that the single-crystal Cu(111) grain reaches the width of the Cu foil (Fig. 1c-e). Further sliding of the whole Cu foil through the central hot zone at the speed of 1.0 cm/min led to a single-crystal Cu(111) foil of $5\times50$ cm$^2$ thus in about 50 minutes (Fig. 1f). The width of the Cu foil is limited by the size of our current furnace tube (that is 5.5 cm in diameter), and the length of the Cu foil is defined by the current foil supplier. In principle, the size of single-crystal Cu(111) foil could be much larger, through appropriate modifications of this approach.

That the Cu(111) foil is a single crystal can be readily seen from optical images. After oxidization in air at 200 ℃ for 5 minutes, the single-crystal region is a homogeneous colour, while the polycrystalline region is of inhomogeneous colour due to surface-index-dependent oxidation levels with clear boundaries (Fig. S1 and Movie S1, online). The surface structure of the 0.5 m-long foil was sampled in many regions by low energy electron diffraction (LEED) measurements, where all the diffraction patterns have three-fold rotational symmetry with identical rotation angles (Fig. 1f-g), which confirms the single-crystal Cu(111) surface structure throughout. The same conclusion can also be drawn from electron backscatter diffraction (EBSD) mapping (Fig. S2, online). In addition, powder and single-crystal X-ray diffraction (XRD) (Fig. S3, online) and transmission electron microscope (TEM) characterizations (Fig. 1h) corroborate that the whole foil has the fcc(111) surface orientation.

In an attempt to understand the mechanism of how the Cu(111) foil is made, Density-Functional-Theory (DFT) calculations and Molecular-Dynamics (MD) simulations were done (see Supplementary Methods for details). The calculated surface energies of three typical low index surfaces (Fig. 2a-c) show that the Cu(111) surface has the lowest formation energy and therefore the transformation from polycrystalline Cu foil to Cu(111) foil is reasonable. Surely the migration of the GBs in the Cu foil requires a driving force. Our MD simulation shows that the Cu atoms near a GB are loosely bonded, and the zone near the GB is in a premelting state at the temperature of 1300 K (Fig. 2d-e). After the GB is premolten, its mobility drastically increases and it can diffuse rapidly in the bulk (Fig. 2e, 6-10 ns). Under a temperature gradient it moves continuously toward the high temperature side of the Cu foil (Fig. 2e, from 11.25 to 30 ns). The speed of GB motion was found to be proportional to the temperature gradient (Fig. 2f) and to increase exponentially with the temperature (Fig. 2g). The simulation shows that the key driving forces of the GB's fast motion and in a preferred direction are the GB pre-melting and the temperature gradient, which agrees well with the classical theory of GB migration[26].

With the very large single-crystal Cu(111) foil, combining our adjacent-oxide-assisted ultrafast graphene growth technique[17] and the principle of the roll-to-roll method[27, 28] (Fig. 3a), in only 20 min a monolayer graphene film with the dimension of $5 \times 50$ cm$^2$ was synthesized on one side of the foil facing the oxide (Fig. 3b-e). The LEED characterizations over the large-area film (graphene: Fig. 3f, Cu: Fig. 3g) show that the graphene film grew epitaxially on the Cu(111) surface. To quantitate the degree of alignment of the graphene islands in a larger scale, 1200 LEED patterns were obtained from randomly chosen positions on the graphene film (Fig. S4 and Movie S2, online), and all of them show nearly identical alignment within <0.5° (Fig. 3h). In batch studies, sub-monolayer growths were done by sliding the foil more quickly through the furnace so that individual islands were present on the foil. Examination of over 3400 individual graphene islands from 165 images (see examples in Fig. S5, online) yielded 2 unaligned islands, suggesting an alignment level of > 99 % (Fig. 3h inset). So, the percentage of 'unalignment' (< 1%) is obviously reduced from previous reports (2-5%)[22, 23]. This almost perfect alignment was further tested by polarized optical images of graphene spin-coated with nematic liquid crystal[29], where the islands of the same orientation show constant optical contrast (Fig. S6, online).

Although the mechanism of the improved island alignment needs future theoretical studies, we believe that it is related to the very flat single-crystal Cu surface and a role of the continuous oxygen supply during graphene growth---the passivation of the active nucleation centres on the Cu foil[11]. If all the disordered nucleation centres (such as impurities) are passivated by oxygen, the graphene nucleation has to be on the clean Cu(111) surface and, therefore, the orientation of the graphene nucleus should be dominated by the Cu(111) surface or the metal steps on the surface. In such a case, the perfect alignment of the graphene nuclei is ensured as previously reported in Ref. [30]. In addition, the oxygen can also suppress the nucleation density[11], accelerate the dehydrogenization reactions and realize the ultrafast growth of graphene[31]. The continuous oxygen supply from the adjacent oxide substrate is the key and marvellous factor to improve the epitaxial graphene growth[32].

It has been stated that the aligned SCG islands can seamlessly merge into a large SCG film during the final stage of graphene film growth to avoid the energetically unfavourable GB formation in graphene[23]. Whether seamless stitching in our procedure has occurred can be tested by both macroscopic and microscopic methods. Hydrogen etching was explored to assess the quality of the stitching between islands[33]. For the graphene obtained from the 'merger' of aligned island edges, no etching lines were observed (Fig. 7Sa-b, online), in contrast to the obvious etching lines between neighbouring unaligned islands (Fig. S7c, online). Large-scale etching with $H_2$ showed that all etched holes are parallel with each other and randomly distributed (Fig. 3i-j and Fig. S7d-i, online), which provides the most direct evidence that there are no rotation angles between different islands. Ultraviolet-light assisted oxidation was conducted to assess the regions where islands could possibly merge. It has been reported that the Cu underlying the grain boundaries is preferentially oxidized[34], and no oxidized 'lines' were observed when the Cu is covered by aligned graphene islands in the regions where they (thus) have merged, demonstrating seamless stitching. (Fig. 3k and Fig. S8, online). Scanning tunnelling microscopy (STM) was also used to evaluate whether the atomic structures in the aligned islands and at the merged regions (Fig. 3l) are identical and defect-free (Fig. 3m-o, see more in Fig. S9, online), and this was found to be the case. In conventional epitaxial growth in surface sciences, the perfect stitching is relatively difficult, presumably due to very hard substrate surface. In the graphene epitaxial growth on Cu(111) surface, the Cu surface is kind of soft (as the growth temperature of 1030 ℃ is close to the Cu melting temperature of 1083 ℃), which greatly facilitates the perfect stitching of aligned graphene domains[24].

Thus, the large SCG islands (~300 μm), nearly perfect alignment (> 99 %) and the seamless stitching of neighbouring islands all ensure the high quality of the synthesized graphene film. To further evaluate the quality, Raman measurements (Fig. 4a) were carried out. The D peak is not observed and the intensity ratio of 2D/G band is homogeneously larger than 2, indicating high-quality monolayer film. Also the scanning electron microscopy (SEM) images demonstrate >99 % monolayer graphene coverage. The measured field effect transistor mobility of the graphene film transferred onto hexagonal boron nitride (h-BN) surface is as high as ~ 23,000 $cm^2V^{-1}s^{-1}$ at 4 K (Fig. 4b and Fig. S10, online) and ~ 15,000 $cm^2V^{-1}s^{-1}$ at room temperature. Prominent quantum Hall plateaus of the graphene film transferred onto a $SiO_2$/Si substrate were observed at 4 K (Fig. 4c). The room temperature sheet resistance of a 2 × 2 $cm^2$ graphene sample, ~ 230 ± 20 Ω/□, is about one-fourth that of our polycrystalline graphene films (Fig. 4d), and is also lower than those in most previous reports for undoped CVD grown graphene films (Table S1-2, online). We note that the measured electrical properties are highly dependent on the transfer process, which may easily bring random impurities and/or strain distribution.[35] Therefore we carried out controlled experiments on samples grown on single-crystal and polycrystalline Cu foils to eliminate the influence of transfer process. Our results do show that SCG films have superior electrical properties to non-SCG films. The above data and its analysis show that the graphene film is of high quality.

The conventional single-crystals are grown from one nucleation centre. The quality of our graphene sample, synthesized via the coalescence of large ultra-highly oriented islands, is very close to that kind of a single crystal although it is undeniable that there may still exist few line defects or point defects that present characterization technique could not detect. This issue can be further clarified in the future if every defect in the superlarge graphene film could be visualized. Also these defects, if existed, can be further diminished by adopting the "etching and regrowth" technique[13].

We suggest that the approach presented here provides a path to the industrial production of large-area high-quality graphene film with all the features of single crystal. With appropriate modifications, the generation of both single-crystal Cu(111) foil and thus single-crystal graphene film up to extremely large scale is thus clearly possible. The method outlined here provides an opportunity for rapid scaling up for industrial-level applications of graphene and seems likely to promote further studies on scaled growth of various other single-crystal 2D materials.

**Conflict of interest**

The authors declare that they have no conflict of interest.

**Acknowledgements**

We thank Feng Wang and Ting Cao for useful discussion and proofreading of the manuscript. This work was supported by National Key R&D Program of China (2016YFA0300903, 2016YFA0300802, 2014CB932500 and 2016YFA0200101), National Natural Science Foundation of China (51522201, 11474006, 11327902, 11234001, 21525310, 91433102 and 21573186), Postdoctoral Innovative Personnel Support Program (BX201700014) and National Program for Thousand Young Talents of China and the Institute for Basic Science (IBS-R019-D1) of Korea.

**Appendix A. Supplementary data**

Supplementary data associated with this article can be found, in the online version, at http://dx.doi.org/***.

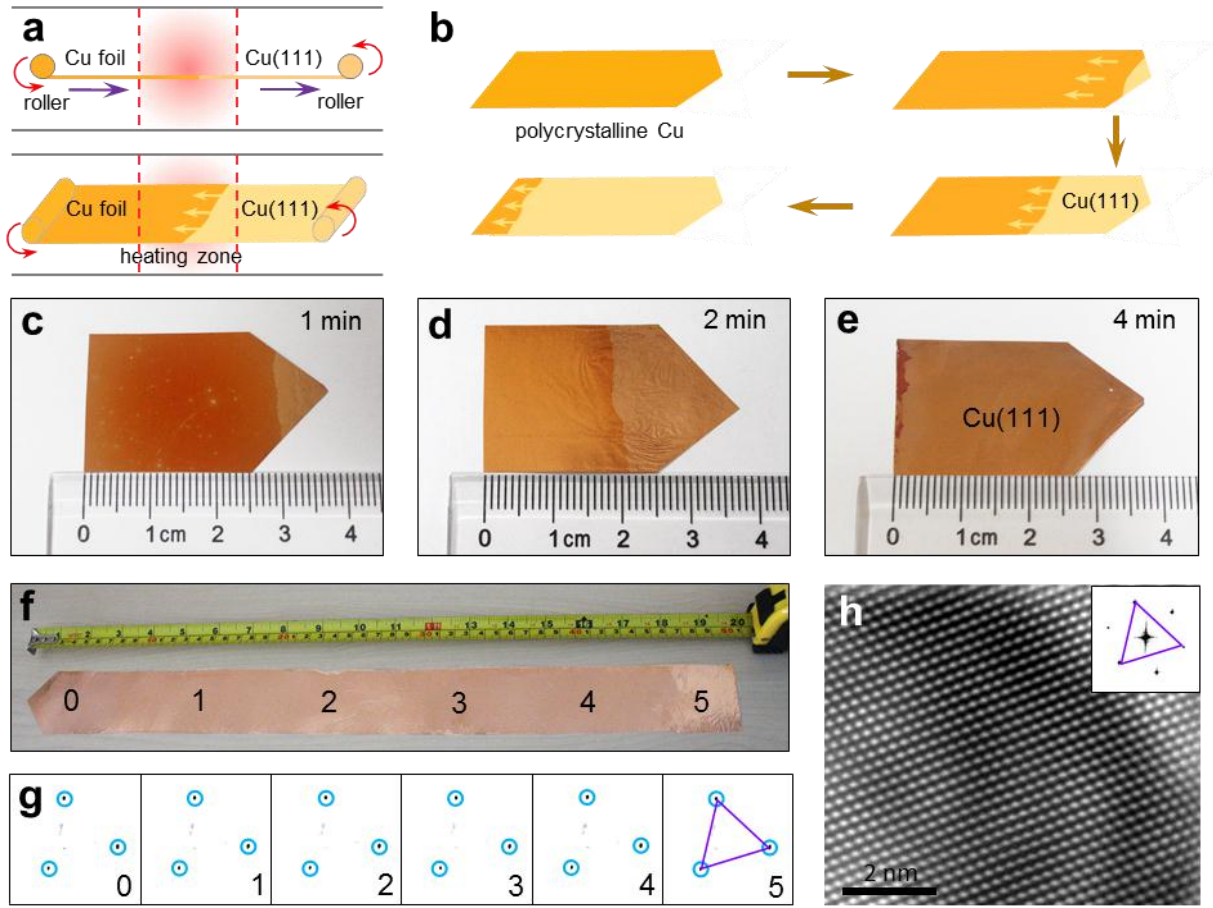

**Fig. 1.** Continuous annealing of polycrystalline Cu foil into single-crystal Cu (111) foil. (a) Schematics of experimental design for the continuous production of single-crystal Cu(111) foil with a hot temperature zone at the central area of the furnace tube. (b) The schematic evolution of the Cu(111) grain growth in the foil during annealing—a single-crystal Cu(111) nucleus formed at the tapered end of the foil and the size of this grain gradually expanded to the full foil width and then the single-crystal Cu(111) region continuously progressed during the continuous sliding of the foil. (c-e) A single-crystal Cu(111) grain is formed and grows larger and larger (photos were taken after 1(c), 2(d) and 4(e) minutes of heating with a central temperature of 1030 °C. (f) The 5×50 $cm^2$ single-crystal Cu(111) foil obtained after ~50 minutes of this process. g) Representative LEED patterns from six different regions of the Cu foil marked in (f). (h) The high-resolution transmission electron microscopy (HRTEM) image of the Cu foil shows its fcc(111) surface orientation. Inset in h: the fast Fourier transformation pattern of the TEM image. The same orientation of the two triangles in (g, h) shows the same crystalline orientation.

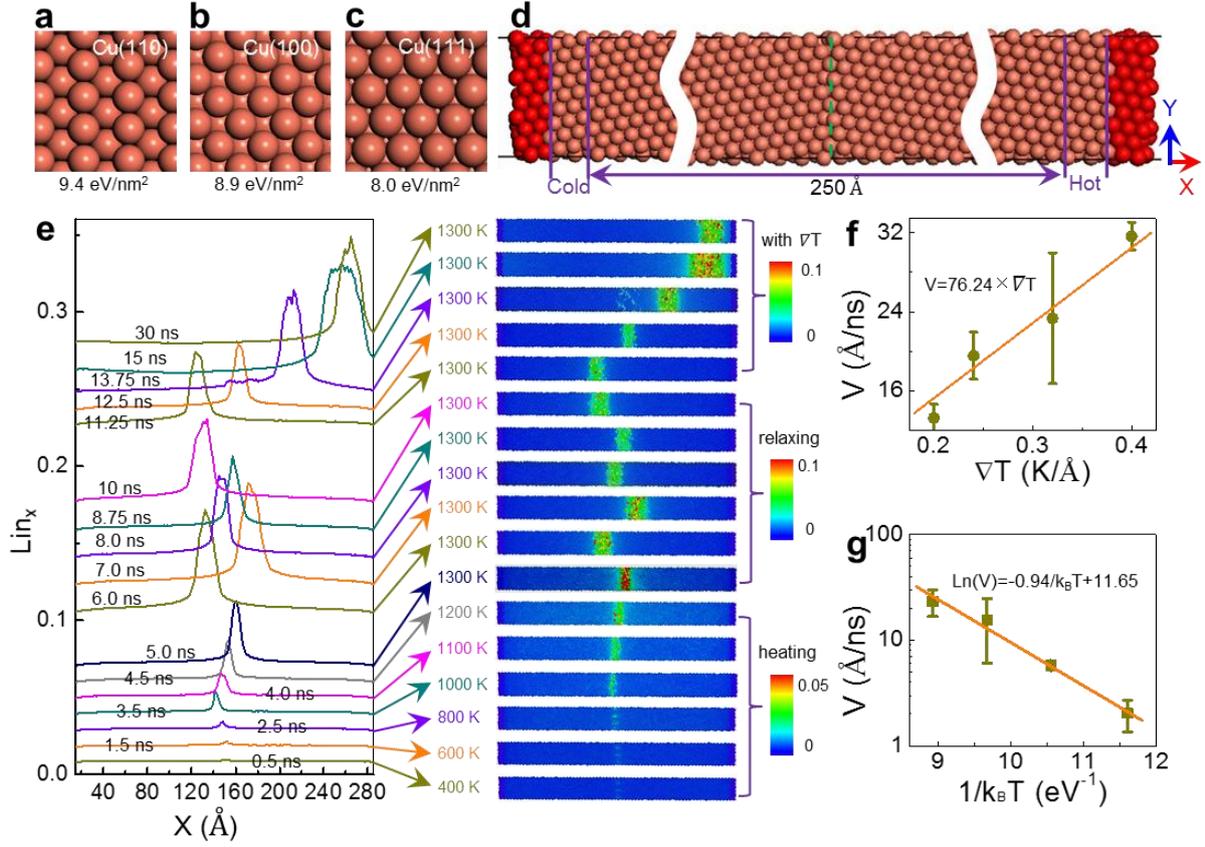

**Fig. 2.** The generation of single-crystal Cu(111) foil as driven by a temperature gradient. (a-c) Structures and formation energies of three typical low index Cu facets, (110), (100) and (111). The (111) surface has the lowest formation energy. (d) A grain boundary (GB) model used for the molecular dynamics (MD) simulation. (e) The averaged Lindemann index along the X axis ($Lin_x$) of the simulated unit cell during the MD simulation (left panel) and the corresponding atomic Lindemann index maps in the simulated unit cell (right panel), where the $Lin_x$ profiles are shifted up for clear visibility. (f) The velocity of the GB migration ($V$) at different temperature gradients $\nabla T$. (G) The velocity of GB migration at different temperatures $T$. The linear fitting lines are shown in (f) and (g).

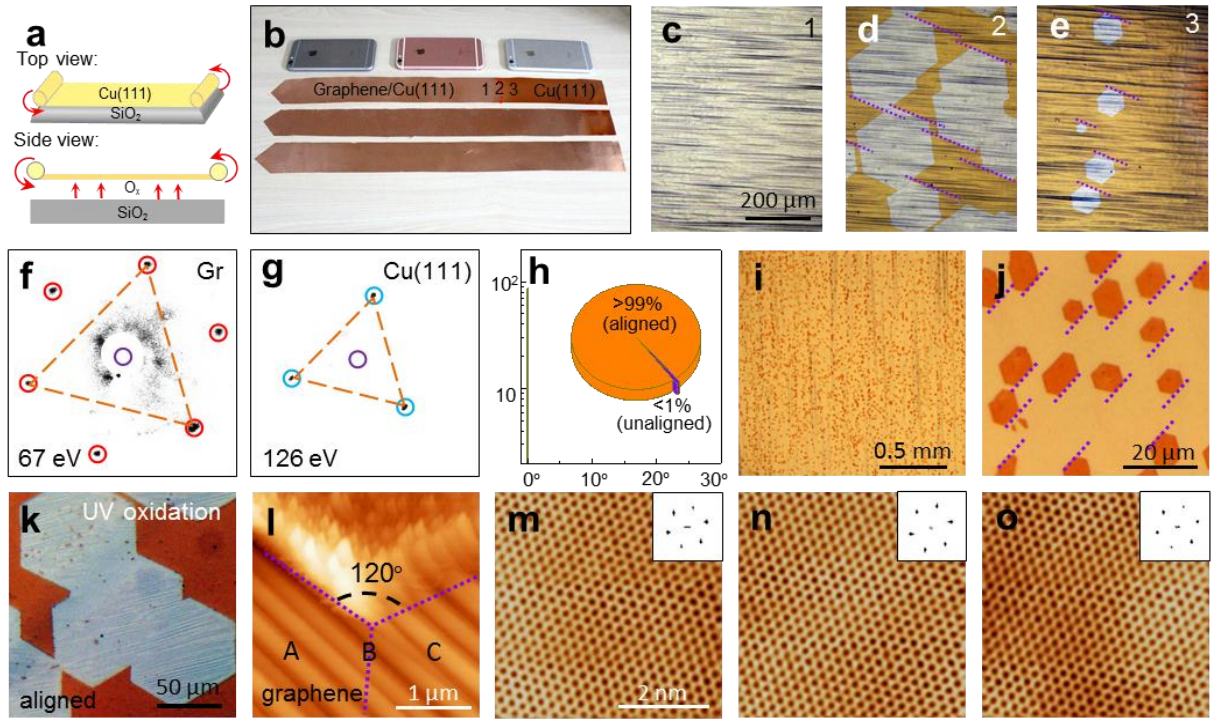

**Fig. 3.** Continuous growth of 5×50 cm² graphene films on the surface of the annealed single-crystal Cu(111) foil. (a) Top and side schematic views of the continuous graphene film growth system, where the Cu(111) foil was placed above a SiO$_2$ substrate with a small separation, for ultrafast growth. (b) Cu(111) foils with graphene coverages of ~ 60% (top), ~ 90% (middle) and 100% (bottom), where the 'shining' parts are graphene/Cu (left side). The three iphone 6s are placed nearby as a reference. (c-e) Optical images of three regions of the graphene covered Cu(111) foil (marked as 1, 2, 3 in b) where graphene fully covered the Cu foil (c), areas with aligned large graphene islands (d) and aligned small graphene islands (e) are clearly seen on the left of, and on the right of, and at the growth front, respectively. (f-g) The LEED patterns of as-grown graphene film (f) and the underlying Cu substrate (g) show that graphene islands grew epitaxially on the Cu(111) surface. (h) The orientation angle distribution of the measured 1200 LEED patterns of the graphene film, showing that all of them are aligned. The inset: percentage of the aligned graphene islands measured from optical images (Fig. S5, online). (i-j) Optical image of the randomly distributed holes formed by H$_2$ etching of the graphene film. Edges of the holes marked by the dashed lines are parallel with each other. (k) Optical images of aligned graphene islands after UV oxidization for 10 min. (l) A large-scale STM image near the corner of two merged aligned graphene islands. (m-o) Representative atomic-resolution STM images corresponding to the A, B, and C regions marked in (l), showing that there are no defects formed during the island coalescence. Insets: fast Fourier transformation patterns of the STM images.

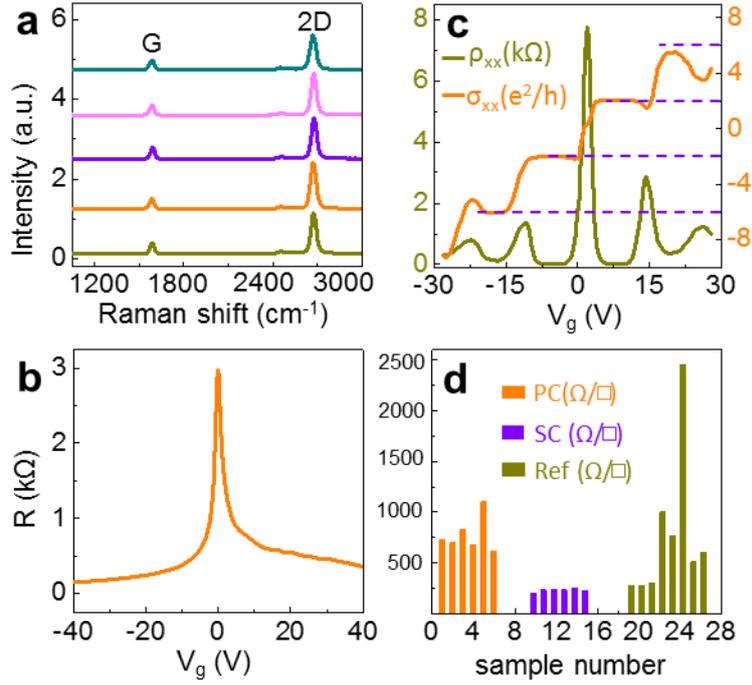

**Fig. 4.** Quality evaluation of graphene film. (a) Representative Raman spectra taken randomly at locations on the whole graphene film. (b) Resistance of graphene versus the back-gate voltage at 4 K. (c) Electrical measurement of graphene Hall bar devices at 4 K under a magnetic field of 9 T. The half-integer plateaus corresponding to filling factor $\upsilon = \pm 2, \pm 6$ are indicated by horizontal lines. (d) Room temperature sheet resistance of a 2 × 2 cm$^2$ sample of polycrystalline (PC) and the single-crystal graphene films and the reference data of undoped graphene films reported previously (Table S1, online).